\documentclass[journal=cmatex,manuscript=article]{achemso}

\usepackage[version=3]{mhchem} % Formula subscripts using \ce{}
\usepackage{pslatex}
\usepackage{xr}
\usepackage{natbib}
\usepackage[dvipsnames]{xcolor} %Used to highlight parts of the text. Remove when submitting to journal (colorbox)

\DeclareGraphicsExtensions{.pdf,.png,.jpg}
\graphicspath{{./Figures/}} % this places all graphics in the figures subdirectory

\usepackage{hyperref}

\makeatletter
\newcommand*{\addFileDependency}[1]{% argument=file name and extension
 \typeout{(#1)}
 \@addtofilelist{#1}
 \IfFileExists{#1}{}{\typeout{No file #1.}}
}
\makeatother

\newcommand*{\myexternaldocument}[1]{%
 \externaldocument{#1}%
 \addFileDependency{#1.tex}%
 \addFileDependency{#1.aux}%
}

\myexternaldocument{Supplemental} %% name of SI file, without .tex extension

\author{Anna A. Berseneva}
\affiliation[ National Renewable Energy Laboratory]
{Materials Science Center, National Laboratory of the Rockies, Golden, CO, 80401, USA}
\altaffiliation{Contributed equally to this work}
\author{Christopher L. Rom}
\affiliation[National Renewable Energy Laboratory]
{Materials Science Center, National Laboratory of the Rockies, Golden, CO, 80401, USA}
\altaffiliation{Contributed equally to this work}
\author{Layton Rudolph}
\affiliation{Department of Chemistry, Colorado State University, Fort Collins, Colorado 80523, USA}
\altaffiliation{Contributed equally to this work}
\author{Yunseung Kuk}
\affiliation{Department of Chemistry, University of Houston, Houston, Texas 77204, USA}
\author{P. Shiv Halasyamani }
\affiliation{Department of Chemistry, University of Houston, Houston, Texas 77204, USA}
\author{Rebecca W. Smaha}
\affiliation[National Renewable Energy Laboratory]
{Materials Science Center, National Laboratory of the Rockies, Golden, CO, 80401, USA}
\author{James R. Neilson}
\affiliation{Department of Chemistry and 
School of Materials Science \& Engineering, Colorado State University, Fort Collins, Colorado 80523, USA}
\author{Andriy Zakutayev}
\affiliation[National Renewable Energy Laboratory]
{Materials Science Center, National Laboratory of the Rockies, Golden, CO, 80401, USA}
\email{Andriy.Zakutayev@nlr.gov}

\title{Synthesis and properties of bulk \ce{Mg3WN4} in a wurtzite-derived structure}

\begin{document}
\begin{abstract}
Experimental synthesis of theoretically predicted materials with controlled elemental coordination environments can lead to realization of useful properties, such as facile ion transport or ferroelectric switching. Among such materials are new ternary nitrides in the Mg-W-N composition space, where several new stable and metastable compounds have been predicted and synthesized recently in bulk and film forms. Here, we report for the first time on the bulk synthesis of \ce{Mg3WN4} in a wurtzite-derived crystal structure via a solid state metathesis reaction. \textit{In situ} synchrotron powder X-ray diffraction shows how the ion exchange proceeds from \ce{Li6WN4 + 3 MgCl2} precursors to \ce{Mg3WN4 + 6 LiCl} products, with the reaction starting slowly near 380 °C and completing by 600 °C, including the presence of a competing disordered rocksalt-derived phase (Mg,W)N above 440 °C. The follow up \textit{ex situ} powder synthesis at 400 °C for 0.5 hour with 10\% excess \ce{MgCl2} reveals the cation-ordered nature of the wurtzite-derived \ce{Mg3WN4} structure with polar symmetry confirmed by second harmonic generation measurements. Optical absorption spectra, chemical composition analysis, and electron microscopy imaging suggests that bulk wurtzite \ce{Mg3WN4} is prone to defect formation. Overall, this study shows that selective \textit{ex situ} synthesis of the phase pure ternary nitrides, informed by \textit{in situ} measurements, is possible by carefully controlling the thermal budget of the reaction, and paves a way towards property characterization of wurtzite \ce{Mg3WN4}.

\end{abstract}

\section{Introduction}

Controlling local elemental coordination environments often leads to beneficial material properties. For example, tetrahedrally-coordinated \ce{Mg^{2+}} is more predisposed towards Mg ionic mobility than \ce{Mg^{2+}} in octahedral environments, according to a prior theoretical study.\cite{rong2015materials, canepa2017high}. However, materials with \ce{Mg^{2+}} in pure tetrahedral environment are difficult to find, since \ce{Mg^{2+}} in octahedral coordination is more common. Another theoretical study \cite{lee2024emerging} aimed at discovery of new tetrahedrally-bonded wurtzite ferroelectric materials beyond (Al,Sc)N \cite{FichtnerAlScN} predicted that \ce{Mg^{2+}} ions in tetrahedral coordination are promising local structural features for facile ferroelectric switching in nitrides with wurtzite-derived crystal structures, compared to much more common tetrahedrally-coordinated \ce{Zn^{2+}} atoms with lower bond ionicities. This makes the hypothetical \ce{Mg3WN4} material a more promising ferroelectric than the previously synthesized \ce{Zn3WN4}.\cite{rom2024low} An additional interesting aspect of the hypothetical \ce{Mg3WN4} is that its reported \ce{Mg3MoN4} cousin shows a collective rather than individual switching pathway, unlike many other wurtzite nitride ferroelectrics. \cite{lee2024emerging}. Thus, the tetrahedrally-coordinated \ce{Mg^{2+}} local structural feature in wurtzite-derived (WZ) \ce{Mg3WN4} makes this material interesting as a potential solid-state Mg ion conductor and as a promising ferroelectric material.

A large computational and experimental study predicted numerous new ternary nitrides and experimentally realized several of them. \cite{sun2019map} Among them, \ce{Mg3WN4} and \ce{Zn3WN4} were predicted in a WZ structure type (NLR MatDB ID 290104) with tetrahedrally coordinated \ce{Mg^{2+}} and \ce{Zn^{2+}} cations.\cite{LanyFERE} According to theoretical calculations, WZ \ce{Zn3WN4} and \ce{Mg3WN4} adopt a polar structure ($Pmn2_1$), with large calculated bandgaps (3.96 eV and 5.17 eV, respectively) based on the GW method \cite{lany2013band, lany2015semiconducting}. As a part of that study, we synthesized \ce{Zn3WN4} in a cation-disordered WZ structure via thin film sputtering \cite{sun2019map}, with optical absorption onset around 2 eV. Subsequently, we reported a cation-ordered polymorph of \ce{Zn3WN4} that exhibited the 2 eV optical absorption onset \cite{rom2024low}. However, experimental synthesis of \ce{Mg3WN4} in a WZ-derived crystal structure has not been reported, and only rocksalt crystal structure is known at this composition. \cite{rom2023bulkMgWN2}

\begin{figure}
 \centering
 \includegraphics[width=3.25in]{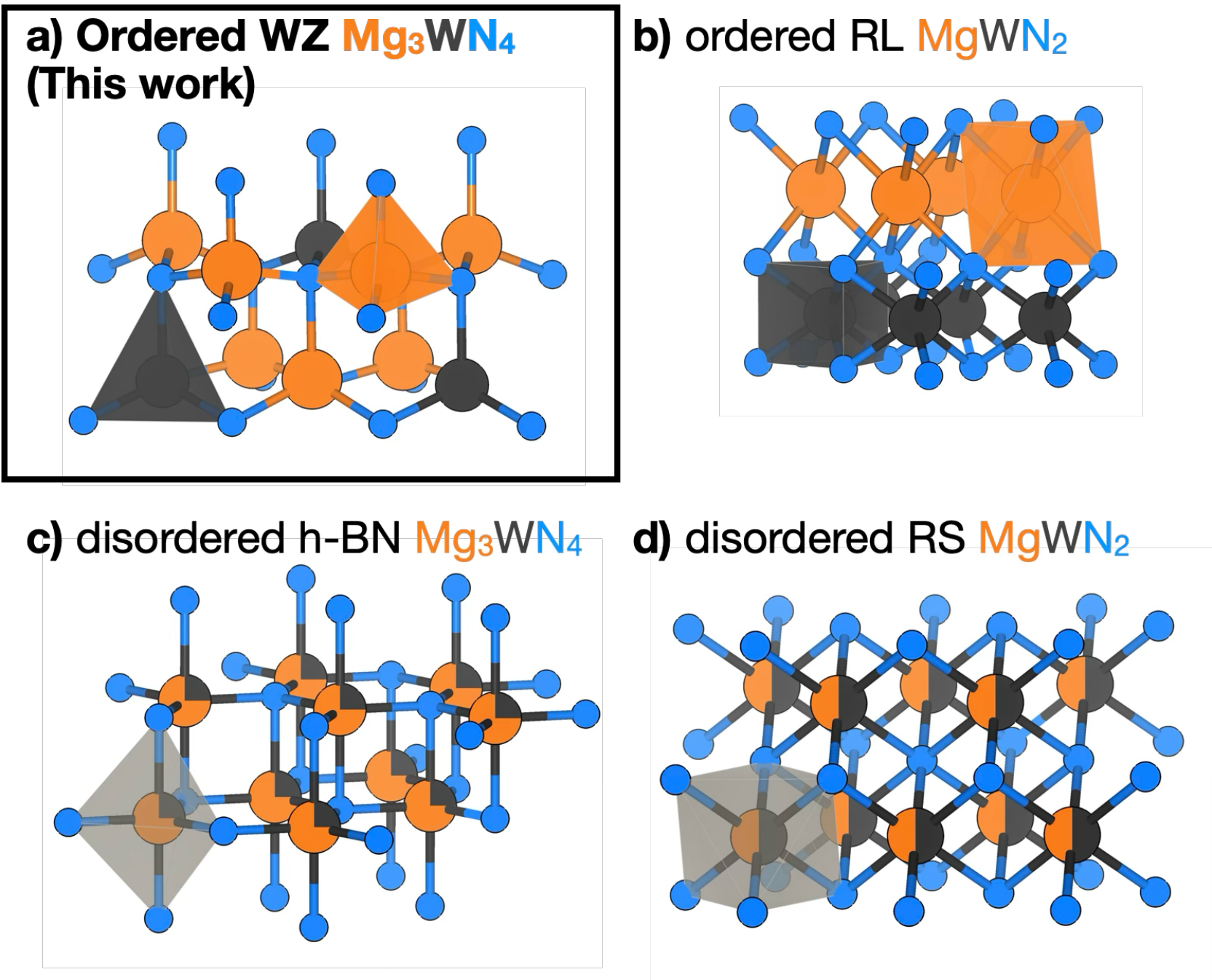}
 \caption{Several structures form in the Mg-W-N phase space including a) cation-ordered, wurtzite-derived \ce{Mg3WN4} (described here), and previously reported b) rockseline, c) h-BN, and d) rocksalt phases.}
 \label{fig:structures}
\end{figure}

Our prior experimental work focusing on the Mg-W-N chemical space realized several new compounds in various structure types.\cite{rom2023bulkMgWN2} We discovered \ce{MgWN2} in a layered 'rockseline' (RL) structure using ceramic synthesis techniques, and used sputtering to yield thin films of \ce{Mg_{$x$}W_{1-$x$}N} in cation-disordered rocksalt (RS; $0.1 < x < 0.9$) and hexagonal boron nitride (h-BN; $0.7 < x < 0.9$) structure types (Figures \ref{fig:structures} and S4). Annealing thin films of $x\sim 0.5$ was used to convert the disordered RS structures to cation-ordered RL structures for \ce{MgWN2},\cite{zakutayev2024synthesis} but for the $x\sim 0.75$ composition \ce{Mg3WN4} remained in the RS structure up until decomposition temperature of 1000 °C.\cite{rom2023bulkMgWN2} Combined with the density functional theory calculations, these experiments showed that RL \ce{MgWN2} defined the convex hull of the Mg-W-N chemical system. Cation-ordered WZ \ce{Mg3WN4} is therefore predicted to be enthalpically metastable at 0K (+0.063 eV per atom above the hull).\cite{rom2023bulkMgWN2}. As a result, the synthesis of WZ \ce{Mg3WN4} has eluded us so far.

More recently, ion exchange approaches have been used to make cation-ordered versions of other ternary nitrides, including metastable structures. Starting from a ternary \ce{Li2ZrN2} precursor, metastable polymorphs of \ce{MgZrN2} and \ce{MgHfN2} with cation-layered structures (space group $R\bar{3}m$) have been synthesized.\cite{rom2025ion} This metastable layered structure contrasts with the calculated ground state cation-ordered structure ($I4_1/amd$) predicted by theory\cite{sun2019map} and with the previously synthesized cation-disordered structure of \ce{MgZrN2}.\cite{rom2021bulk, todd2021twostep, bauers2019ternaryrocksaltsemiconductors} During this cation-exchange reaction, \ce{2Li^{+}} and \ce{Mg^{2+}} undergo ion exchange topochemically, preserving the layers of octahedral [\ce{ZrN6}] according to \textit{in situ} synchrotron powder X-ray diffraction (PXRD) measurements. In the past, similar cation-exchange reactions have been used to make other metastable layered nitrides, such as \ce{CuTaN2}\cite{yang2013strongCuTaN2} and \ce{CuNbN2}.\cite{zakutayev2014experimentalCuNbN2} Together, these results indicate that ion exchange approaches may be suitable for experimentally realizing the theoretically predicted metastable \ce{Mg3WN4} with tetrahedral \ce{Mg^{2+}} coordination and promising ferroelectric switching or Mg ion transport properties.

Here, we report the synthesis of \ce{Mg3WN4} in a cation-ordered WZ structure ($Pmn2_1$), as observed by \textit{in situ} and verified by \textit{ex situ} PXRD data. The \ce{Mg3WN4} material forms from the reaction of \ce{Li6WN4} and \ce{MgCl2} precursors in the range of 400--600 °C. The results of the \textit{ex situ} PXRD measurements indicate that this cation-ordered WZ phase forms alongside a cation-disordered RS impurity. \textit{In situ} synchrotron measurements show that the WZ \ce{Mg3WN4} phase begins forming at lower temperatures (ca. 380 °C) than the RS polymorph (ca. 440 °C). This knowledge let us synthesize the phase-pure WZ \ce{Mg3WN4} compound at 400 °C for 0.5 hour with 10\% excess in the lab by washing away unreacted \ce{Li6WN4}, which is still present at these lower temperatures. Obtaining a phase pure product allows us to measure optical absorption spectra with diffuse reflectance (DR) spectroscopy, comparing \ce{Mg3WN4} to the \ce{Zn3WN4} analog, as well as to perform second harmonic generation (SHG) measurements to support the polar space group ($Pmn2_1$) of both of these materials. This work, informed by \textit{in situ} measurements, establishes the \textit{ex situ} synthesizability of phase-pure \ce{Mg3WN4} in a cation-ordered wurtzite-derived structure, which is of significant interest for possible ferroelectric switching or Mg ionic conductivity applications.

\section{Methods}
\subsection{Synthesis }
As some precursors are highly moisture sensitive, all precursors were prepared and stored in an argon-filled glovebox (\ce{O2} $< 0.1$~ppm, \ce{H2O} $< 0.1$~ppm) unless explicitly mentioned. 
\ce{Li3N} ($\geq99.5$\%, 80 mesh, Sigma Aldrich), W (99.95\%, $<1$ micron powder, Thermofisher Scientific), and \ce{MgCl2} (Sigma-Aldrich, 99.99\%, AnhydroBeads) were used as received.

\ce{Zn3WN4} was synthesized using the method from our previous report.\cite{rom2024low} \ce{Li6WN4} was synthesized using a method modified from literature, \cite{yuan2005synthesisLi6WN4} as we described previously.\cite{rom2024low}
Briefly, solid precursors (\ce{2.1 Li3N + W}, where the ca. 5 mol\% excess \ce{Li3N} accounts for loss by evaporation) were ground with an agate mortar and pestle and loaded into custom Zr crucibles with Zr lids (ca. 1 g loose powder), and then heated under flowing \ce{N2} (50 sccm, 99.999\% purity) with a 5 °C/min ramp followed by a 12 h dwell at 850 °C. Samples were cooled by turning off the furnace, then recovered into glovebox without air exposure. The air-free transfer of reagents and final product was done with a tube equipped with the leak-free quick disconnect flange sets by the end of the tube. The resulting powders were beige in color and phase pure by PXRD.

Metathesis reactions were performed by homogenizing \ce{3 MgCl2 + Li6WN4} with an agate mortar and pestle, pelletizing at ca. 1 ton pressure in a 6 mm die, loading the pellet into an alumina crucible, and heating without exposure to air, as described in the text. Post-reaction, samples were recovered into the glovebox without air exposure. The byproduct LiCl was removed via washing with anhydrous methanol in the glovebox. 

The purest sample was obtained via reaction between \ce{3.3 MgCl2} and \ce{Li6WN4}. Solid precursors were ground, pressed into a pellet, loaded into an alumina crucible, and then heated under flowing \ce{N2} (50 sccm, 99.999\% purity) with a 10 °C/min ramp followed by a 0.5 h dwell at 400 °C. Sample was cooled by turning off the furnace, recovered into glovebox without air exposure, then washed with anhydrous methanol.

\subsection{Compositional Analysis}
Scanning electron microscopy (SEM) and energy-dispersive spectroscopy (EDS) were performed to visualize particles and semi-quantitatively assess composition. 
SEM imaging was done on a Hitachi S-4800 SEM operating at a 15 keV accelerating voltage and 10 $\mu$A beam current. 
Elemental analysis was conducted by EDS on the same instrument using the included Pathfinder analysis software for quantification. 
EDS was performed directly on powder mounted on an SEM stub with carbon tape. 
Spectra were acquired for 60 s. EDS mapping was performed by acquiring data for 6 minutes.
Note that C, Al, and Si are always present in the EDS spectra coming from carbon tape, the Al SEM stub, and the detector, respectively.

Wavelength dispersive X-ray fluorescence (WD-XRF) was performed to semi-quantitatively assess composition using a Rigaku ZSX PrimusIV. Data collection was performed on powder capped with the Enthom 3.0 µm thin film in the Ar glovebox. 
Note that O element is always present in this set up coming from the Enthom film and that N element cannot be determined in this configuration due to the full N K$\alpha$ edge absorption by the film.
\subsection{\textit{Ex situ} laboratory PXRD}
The products of all reactions were characterized by powder X-ray diffraction (PXRD). 
Laboratory PXRD patterns were collected on a Rigaku Ultima IV diffractometer and a Rigaku SmartLab in Bragg-Brentano geometry with Cu K\textsubscript{$\alpha$} X-ray radiation at room temperature. For Rigaku Ultima, we used a 10 mm slit, a K\textsubscript{$\beta$} filter, and a 0.02° step with a 0.5 hour collection, several scans for the same samples were merged for the Rietveld refinement. For Rigaku SmartLab, we used a K\textsubscript{$\beta$} filter, and a 0.001° step size with a 1 hour collection; the scan was collected with a variable slit, therefore slit correction was applied in the SmartLab Studio II software. 
All reaction products were initially prepared for PXRD measurements inside the glovebox. Powder was placed on off-axis cut silicon single crystal wafers to reduce background scattering and then covered with polyimide tape to impede exposure to atmosphere. 
After \ce{Mg3WN4} was determined to be moderately air stable, PXRD patterns of the washed \ce{Mg3WN4} were collected without polyimide tape to decrease the background signal.
Rietveld refinements were performed using TOPAS-64 v6 \cite{coelho2018topas} and GSAS-II.\cite{GSAS} Lattice parameters, atom coordinates, microstrain broadening, and thermal parameters were refined for the the purest WZ \ce{Mg3WN4} phase (\textit{Pmn}2\textsubscript{1},
\textit{R}\textsubscript{wp} = 2.863 \%) and presented in Table \ref{tab:rr_results_wzpure}. For a stable refinement, \textit{U}\textsubscript{iso} parameters for N atoms were fixed at 0.01~\AA$^2$ and not allowed to refine; thermal parameters for the Mg and W sites were freely modeled, with \textit{U}\textsubscript{iso} for Mg1 and Mg2 constrained to be equivalent. Free refinement of metal occupancies resulted in values close to 1 for Mg1 and W1, while Mg2 refined to 0.95. Addition of a RS \ce{MgWN2} phase led to 5 wt. \% of RS phase. Both free metal occupancy refinement and addition of a RS phase insignificantly affected \textit{R}\textsubscript{wp} and therefore neither was used in the final model. More information on the refinement models and constraint choices are described in Supporting Information.

\subsection{\textit{In situ} synchrotron PXRD}
\textit{In situ} synchrotron PXRD measurements were conducted at beamline 17-BM-B of the Advanced Photon Source at Argonne National Laboratory. For these experiments ($\lambda = $0.25306 Å), the PerkinElmer plate detector was positioned 800~mm away from the sample.
Homogenized precursors were packed into quartz capillaries in an Ar glovebox and flame-sealed under vacuum ($<40$ mTorr). 
Capillaries were loaded into a flow-cell apparatus\cite{chupas2008versatile} and heated at 10~°C/min to the specified temperature. 
A thermocouple was placed against the tip of the sample capillary, approximately 2 mm horizontally from the position of the X-ray beam. 
Diffraction pattern images were collected every 30~s by summing 100 exposures of 0.1 s each (10 s of summed exposure), followed by 20 s of deadtime. 
Images collected from the plate detector were radially integrated using GSAS-II and using a silicon standard. 

\begin{figure}[ht!]
 \centering
 \includegraphics[width=3.25in]{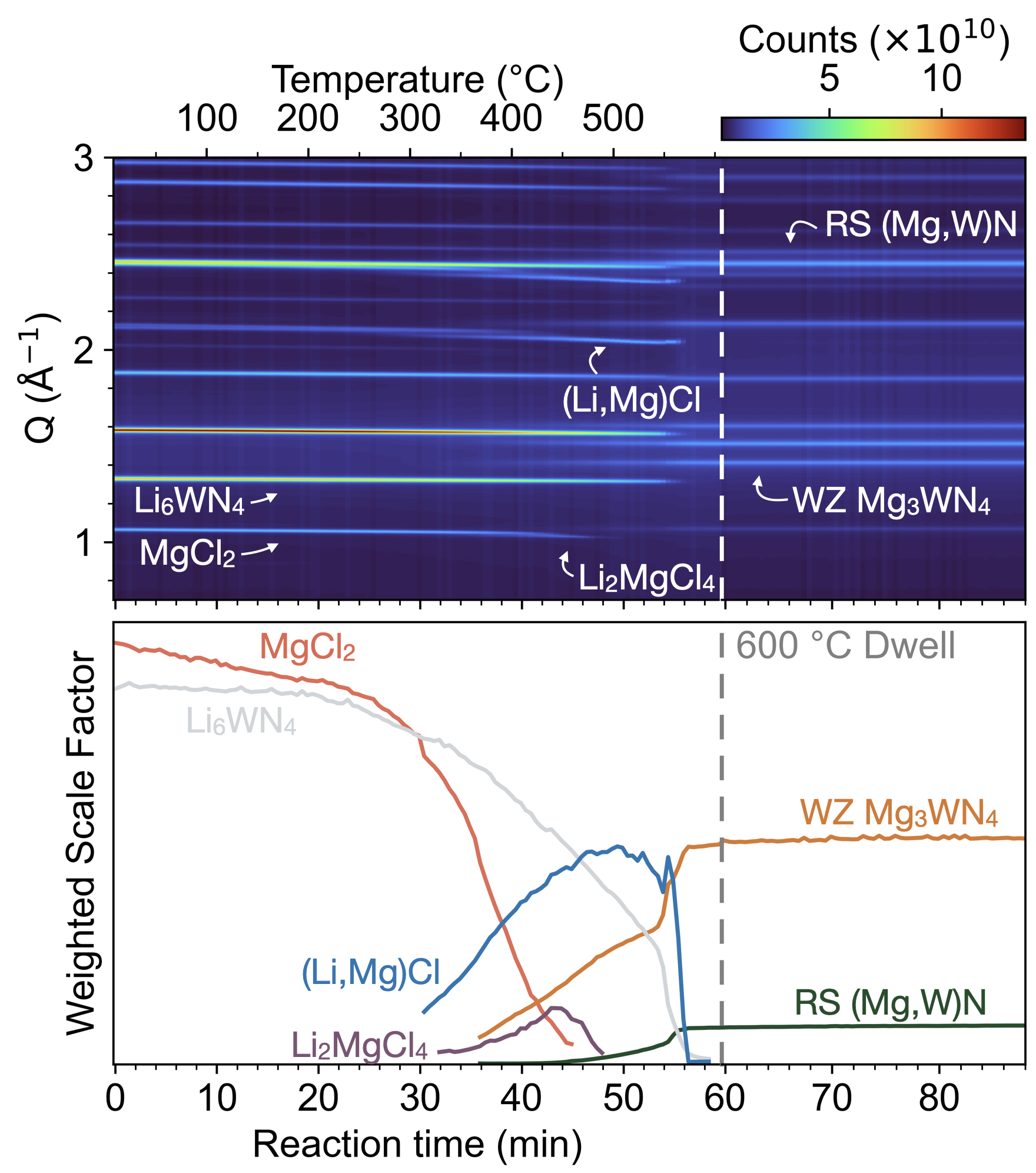}
 \caption{\textit{In situ} synchrotron PXRD of the reaction between \ce{3MgCl2 + Li6WN4}, showing that the reactivity starts close to 400 °C and that the reaction is complete by 600 °C. We represent the \ce{Li2MgCl4} phase using the normalized peak intensity at $Q$ = 1.026191 Å$^{-1}$ as a proxy for the weighted scale factor (WSF).}
 \label{fig:insitu_xrd}
 % Source:
\end{figure}

Sequential Rietveld refinements were conducted on \textit{in situ} synchrotron PXRD datasets using TOPAS-64 v6.\cite{coelho2018topas} 
Lattice parameters, background terms, and scale factors were refined for each phase as a function of temperature, while atomic coordinates and occupancies were held constant at the initial values of the reference structure. 
A weighted scale factor (WSF) was calculated for each phase $p$ as a product of scale factor $S$, cell volume $V$, and cell mass $M$: $(WSF)_p = S_p•V_p•M_p$.\cite{todd2019yttrium}
To derive a mole value for each phase we used $(WSF)_p / M_rp$ formula where $M_rp$ is molar mass and normalized it for all phases with the initial value for \ce{Li6WN4} as 1 (Figures S2 and S3). Data presented in mole scale factor supports the inclusion of Mg into LiCl phase and/or \ce{Li2MgCl4} formation. Owing to substantial overlap between the RS (Li,Mg)Cl phase and \ce{Li2MgCl4}, we were not able to reliably extract the WSF for both phases while allowing the unit cell size to refine. 
Instead, we represent the \ce{Li2MgCl4} phase in Figure \ref{fig:insitu_xrd} using the normalized peak intensity at $Q = 1.026191$~Å$^{-1}$ as a proxy for its WSF. 
We note that amorphous and liquid phases are inherently not observed in powder diffraction measurements and therefore cannot be accurately included in this analysis. 
A Lorentzian size broadening term was refined for each phase to model the peak shape using the pattern showing the greatest intensity of the relevant phase; this term was then fixed for the sequential refinements to better account for changes in intensity. 
To improve reliability in the sequential refinement, isotropic displacement parameters ($B_{iso}$) were fixed at $1$~\AA$^2$ for all atoms, but we note that this is likely not physical for a variable temperature investigation.

\subsection{Optical characterization}
The absorption of the bulk \ce{Mg3WN4} and \ce{Zn3WN4} samples was measured using a Cary 7000 optical spectrophotometer with the sample mounted outside an integrating sphere, diffuse reflectance accessory (DRA) module, using a 1/4”-diameter aperture plate attached to the reflectance port. Ground powder samples were pressed in between two 1/16”-thick fused silica glass slides (1”×1”) from GM Quartz and fixed with epoxy applied around the perimeter of the slides. After baselining with the glass slides, the absorption spectra were collected in the 190--800 nm range with reduced slit, with 120 nm/min rate and 1 nm step size.

Powder second-harmonic generation (SHG) measurements were carried out using a modified Kurtz–Perry powder technique,\cite{KTmethod} employing an Nd:YAG laser operating at 1064 nm and a Ho:YAG laser at 2.09 µm as fundamental light sources. KH\textsubscript{2}PO\textsubscript{4} and AgGaS\textsubscript{2} were used as reference materials for SHG measurements at 1064 nm and 2.09 µm, respectively. Polycrystalline samples of \ce{Mg3WN4}, \ce{Zn3WN4}, and the reference materials were packed into sample holders. The SHG signals were collected using a photomultiplier tube and recorded with an oscilloscope.

\section{Results and Discussion}
We initially synthesized \ce{Mg3WN4} powders via a metathesis reaction at ambient pressure in flowing nitrogen gas:
\begin{equation}
 \ce{Li6WN4 + 3 MgCl2 \rightarrow Mg3WN4 + 6 LiCl}
 \label{eq:main_rxn}
\end{equation}
The un-optimized reaction of heating at 600 °C for 12 h yields a black powder.
Washing with anhydrous methanol removes the LiCl byproduct. 
The EDX in SEM quantifies atomic percentages as shown in Table S4, though the N content could not be determined precisely. The Mg/W ratio of $\sim$3.5--4 indicates that \ce{Mg3WN4} is the likely product. These composition data are derived from the EDS spectra collected from the three separate agglomerates of crystallites shown in Figure S12. Moreover, the cation composition of the bulk sample was confirmed by wavelength dispersive X-ray fluorescence (WD-XRF), matching well with the EDS results (Table S4).

\textit{Ex situ} laboratory PXRD data on the washed reaction products show that \ce{Mg3WN4} phase is the main product (close to 70 wt. \%), in the orthorhombic space group $Pmn2_1$ (no. 31). This structure is derived from the polar wurtzite (WZ) crystal structure by ordering of \ce{Mg^{2+}} and \ce{W^{6+}} on cation sublattice, and hence is also polar. \cite{wurtzites} Other possible cation-ordered derivatives of the WZ structure like $P31c$ (no. 159) \cite{LanyFERE}, or the potential non-polar h-BN or \ce{BeO2} parent compounds, do not fit the PXRD data, as shown in Figure S4. In addition to the WZ phase, a substantial amount of disordered RS phase (close to 30 wt. \%) formed as well (Figure S5).
As the cation composition of this RS phase is unknown, we call it ''RS (Mg,W)N,'' although Rietveld analysis shows that \ce{Mg_{0.5}W_{0.5}N} fits the PXRD data reasonably well. 
Refined parameters for these two phases are shown in Tables S2 and S3. 
Both nitride phases are stable in air (Figure S9). 

\textit{In situ} synchrotron PXRD shows that the reaction between \ce{3 MgCl2 + Li6WN4} has a small temperature window between 380 °C and 440 °C for the selective synthesis of WZ \ce{Mg3WN4} over RS (Mg,W)N (Figure \ref{fig:insitu_xrd}). 
Reactivity is first observed at 280 °C, as the precursor peaks decrease in intensity and (Li,Mg)Cl peaks appear. 
By 380 °C, WZ \ce{Mg3WN4} peaks appear but are weak and grow slowly. 
Near 440 °C, peaks consistent with RS (Mg,W)N grow in.
In the 410 °C to 510 °C range, \ce{MgCl2} peaks fade out since it is consumed for nitride formation and reaction with LiCl: \ce{2LiCl + MgCl2 -> Li2MgCl4}. 
At 510 °C, the spinel \ce{Li2MgCl4} phase converges with the RS (Li,Mg)Cl phase. 
These changes are consistent with the LiCl-\ce{MgCl2} phase diagram.\cite{lutz1979kenntnis}
Slightly before the (Li,Mg)Cl chloride melts (540 °C), the \ce{Li6WN4} precursor is completely consumed and both sets of nitride peaks grow rapidly. 
By 580 °C, WZ \ce{Mg3WN4} and RS (Mg,W)N are the only crystalline phases present (in a flux of liquid LiCl), and no further reaction is observed. 
No other peaks appear or disappear during the 30 min dwell, nor do the nitride peaks change in intensity.

\begin{figure}[ht!]
 \centering
 \includegraphics[width=6in]{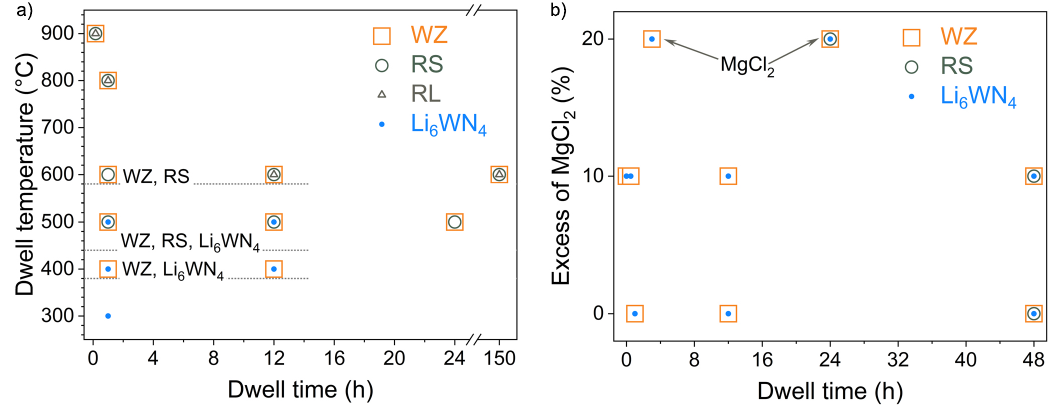}
 \caption{Synthesis map for the \ce{3MgCl2 + Li6WN4} mixture (a) and the (3+x)\ce{MgCl2 + Li6WN4} mixture at 400 °C (b). Symbols represent \textit{ex situ} data and dotted lines correspond to the temperatures of phase transformations from \textit{in situ} measurements.}
 \label{fig:synmap}
\end{figure}
Following \textit{in situ} PXRD, we attempted several synthetic conditions \textit{ex situ} to isolate phase pure WZ \ce{Mg3WN4}. These experimental results are shown by the synthesis map for the \ce{3MgCl2 + Li6WN4} mixture with the dwell time and temperature as the map coordinates (Figure \ref{fig:synmap}a). Because all PXRD data for the map were collected and analysed for unwashed samples, we did not estimate the ratio of products. For clarity, only W-containing phases are shown in Figure \ref{fig:synmap}, since every unwashed product also contains LiCl-salts. At a low dwell time ($<$ 24 h), all our \textit{ex situ} experimental results align well with the \textit{in situ} synchrotron data, shown as a dotted line in Figure \ref{fig:synmap}a. The syntheses at $\leq$ 500 °C showed residual unreacted \ce{Li6WN4} (Figure S6), while prolonged heating ($>$ 24 h) at $\geq$ 500 °C leads to complete consumption of \ce{Li6WN4}. Interestingly, heating at 600 °C for up to 150 h did not change the WZ \ce{Mg3WN4} structure (Figure S7). This result indicates that even though WZ is enthalpically metastable at 0 K \cite{rom2023bulkMgWN2}, it is stable enough to withstand 600 °C for 150 hours, which is encouraging for practical applications of this material. At 800 °C and above, the \ce{MgWN2} rockseline (RL) structure can be detected (Figure S6), indicating decomposition via Mg species gas phase loss. Even as low as 600 °C, trace amounts of \ce{MgWN2} (less than 1 wt. \% is visible only for washed sample) appear in PXRD after heating for 12 h (Figure S5). 

\begin{figure}[ht!]
 \centering
 \includegraphics[width=7in]{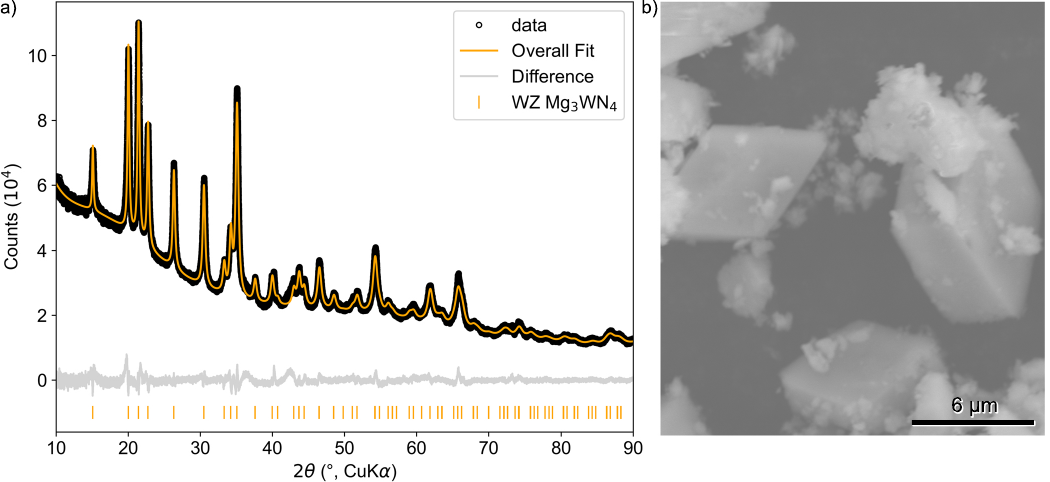}
 \caption{Refinement of {ex situ} laboratory PXRD data (a) and SEM image (b) of WZ \ce{Mg3WN4} synthesized at 400 °C for 0.5 h (left). }
 \label{fig:wz}
\end{figure}

Since the RS phase formation was observed in \textit{in situ} studies above 440 °C, we explored the synthesis map for \ce{MgCl2-Li6WN4} system more thoroughly at 400 °C (Figure \ref{fig:synmap}b). The increased dwell time at 400 °C led to the appearance of the RS \ce{(Mg, W)N} phase, whose crystallinity improved with increased reaction time (24 vs. 48 hours, Figure S10). One of the possible reasons for the challenges associated with WZ \ce{Mg3WN4} phase purity was that formation of (Li,Mg)Cl and \ce{Li2MgCl4} creates a Mg-poor reaction media.\cite{ref1, rom2024mechanistically} However, adding excess \ce{MgCl2} to the reaction at 10\% and 20\% did not affect reaction outcome, i.e., unreacted \ce{Li6WN4} (Figures \ref{fig:synmap} and S11). More interestingly, there is a noticeable amount of unreacted \ce{MgCl2} salt present for the 20\%-excess \ce{MgCl2} reactions (Figure S11). Therefore, to eliminate the \ce{Li6WN4} impurity, we employed methanol washing to dissolve/decompose the side product, leaving only WZ \ce{Mg3WN4} as the final powder (Figure \ref{fig:wz}). The purest WZ \ce{Mg3WN4} was obtained in the reaction with 10\%-excess \ce{MgCl2}, heated at 400 °C for 0.5 h.

\begin{table}[]
\begin{tabular}{llllllll}
site & element & mult. & x  & y  & z  & occupancy & $U_\textrm{iso}$\\ \hline
Mg1 & Mg & 4 & 0.23713(26) & 0.1517(15) & 0.1790(18)& 1& 0.0001(8)\\
Mg2 & Mg & 2 & 0  & 0.3383(18) & 0.6722(16)& 1& 0.0001(8)\\
W1 & W & 2 & 0  & 0.67043(26) & 0.1859(4)& 1& 0.01050(10)\\
N1 & N & 4 & 0.2539(7) & 0.8100(12)& 0.0388(6)& 1  & 0.01\\
N2 & N & 2 & 0  & 0.3820(13) & 0.1028(12)& 1  & 0.01\\
N3 & N & 2 & 0  & 0.7074(13) & 0.5385(11)& 1& 0.01\\\hline
\end{tabular}
 \caption{Atomic coordinates for WZ \ce{Mg3WN4} synthesized at 400 °C for 0.5 h from Rietveld refinements of laboratory PXRD data in orthorhombic space group $Pmn2_1$ (no. 31). Cell parameters refined to $a = 6.75489(16)$, $b = 5.85005(13)$, and $c = 5.22891(10)$ Å. The sample was obtained by heating at 400 °C for 0.5 hour. }
 
 \label{tab:rr_results_wzpure}
\end{table}

Rietveld refinement of the WZ \ce{Mg3WN4} synthesized via heating at 400 °C for 0.5 hour resulted in a very similar structure to that obtained during initial synthesis at 600 °C (Figures \ref{fig:wz}a and S5 and Tables \ref{tab:rr_results_wzpure} and S2). The main difference between these structures is that WZ \ce{Mg3WN4} at 400 °C does not exhibit anti-site mixing as at 600 °C. WZ \ce{Mg3WN4} structure obtained at 600 °C has cation mixing on the Mg sites with the mixing level ranging at 3-4 \% (Table S2). At the same time, the disordered refinement model is not preferential in case of WZ \ce{Mg3WN4} synthesized at 400 °C, and free refinement only suggests electron deficiency on Mg2 site which cannot be modeled with heavier W atom. 

\begin{figure}[ht!]
 \centering
 \includegraphics[width=7 in]{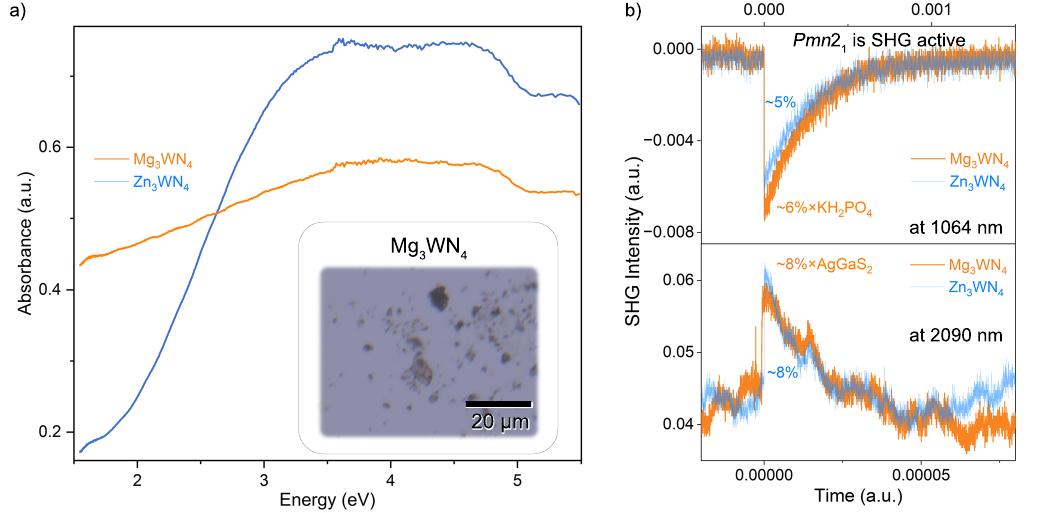}
 \caption{Absorbance spectra (a) and SHG activity (b) of WZ \ce{Mg3WN4} and \ce{Zn3WN4}. The inset shows optical image of WZ \ce{Mg3WN4} sample.}
 \label{fig:dr}
 % Source:
\end{figure}

SEM images of the WZ \ce{Mg3WN4} synthesized via heating at 400 °C for 0.5 hour demonstrated diamond-shaped 5×5×2 $\mu m^3$ single crystal particles with some dark amorphous material covering their surface (Figure \ref{fig:wz}b). In optical microscope, those particles appeared to be transparent colorless crystallites covered with amorphous black powder (Figure \ref{fig:dr}a inset). EDS mapping showed homogeneous Mg and W distribution for both particles and amorphous material (Figure S13). Estimating the Mg/W ratio for particles resulted in 2:1 (Areas 1 and 2 in in Figure S16), suggesting a new material composition \ce{Li2Mg2WN4} with WZ-derived crystal structure, which has not been theoretically or experimentally reported to date. For amorphous material, the Mg/W ratio was higher (2.6:1) with increased chlorine content, which might be salt residue or small fraction of some other amorphous material undetected by XRD (Figure S14 and Table S5). As we increased reaction time from 0.5 to 12 hours at 400 °C, we noticed the disintegration of the diamond-shape crystallites (Area 1 in Figure S15) and the formation of agglomerates consisting of smaller particles (Areas 2 and 3 in Figure S15). EDS analysis of those three areas resulted in 2.8:1 Mg/W ratio, suggesting that more Mg intercalation in the WZ \ce{Mg3WN4} structure destroys the diamond-shaped morphology, also altering color of the sample towards gray or black.

Similarly, WZ \ce{Mg3WN4} powder prepared at 600 °C for 12 hours resulted in even more substantial agglomerates forming (Figure S16). In this case, diamond-shaped WZ \ce{Mg3WN4} particles with the 3.1:1 Mg/W ratio are barely preserved (Area 1 in Figure S16). Even though these EDS results are quantitative and not qualitative, due to the large electron difference between Mg and W affecting results, the measured differences suggest that there is some Mg/W off-stoichiometry ranging between 2 to 3 that can be stabilized in the WZ \ce{Mg3WN4} structure at various synthetic conditions. A smaller Mg/W ratio can be tentatively explained with Li or O presence in the WZ \ce{Mg3WN4} structure, while a higher Mg/W ratio might be associated with Mg substitution on the W site. 

We collected diffuse reflectance spectra for the WZ \ce{Mg3WN4} sample and its Zn-analog.\cite{rom2024low} While WZ \ce{Zn3WN4} has a well defined optical absorption onset above 2 eV, WZ \ce{Mg3WN4} exhibits a much slower onset with absorption slowly increasing from $<$1.5 eV to $>$4 eV (the data collection limit, Figure \ref{fig:dr}a). It is likely that this optical absorption in the visible region of the spectrum correspond to defective WZ structure masking the intrinsic band gap of WZ \ce{Mg3WN4}. Defects like Mg/W anti-site disorder and Li or O substitution in the structure can affect the density of states between the conduction or valence band, resulting in gray or even black color, even though WZ \ce{Mg3WN4} is predicted to be a wide-bandgap semiconductors with expected yellow or white color. Therefore, defects in the WZ \ce{Mg3WN4} structure play a crucial role in material properties and should be understood and mitigated before characterization of ferroelectricity or \ce{Mg^{2+}} ion conductivity in this material.

Nevertheless, to probe the material properties related to ferroelectric applications, we performed second harmonic generation (SHG) measurement for WZ \ce{Mg3WN4} reported here, as well as for its Zn-analog---WZ \ce{Zn3WN4}.\cite{rom2024low} The SHG intensities of Mg\textsubscript{3}WN\textsubscript{4} and Zn\textsubscript{3}WN\textsubscript{4} were found to be approximately 0.08 times that of AgGaS\textsubscript{2} at 2090 nm (Figure \ref{fig:dr}b). At 1064 nm, the SHG responses of WZ \ce{Mg3WN4} and \ce{Zn3WN4} correspond to approximately 0.06 and 0.05 times that of KH\textsubscript{2}PO\textsubscript{4}, respectively (Figure \ref{fig:dr}b). Although the observed SHG signals are relatively weak, these results provide clear evidence for non-centrosymmetric structure of WZ \ce{Mg3WN4} and \ce{Zn3WN4} materials, and support their crystallization in a polar $Pmn2_1$ space group determined from XRD.

\section{Conclusion}
This paper details the successful synthesis of the cation-ordered wurtzite structure of \ce{Mg3WN4} via a metathesis reaction \ce{3 MgCl2 + Li6WN4}. \textit{In situ} synchrotron PXRD shows that a main contaminant, RS (Mg,W)N, begins forming at a slightly higher temperature than WZ \ce{Mg3WN4} (440 °C vs. 380 °C). Narrowing the temperature window to 400 °C for \textit{ex situ} WZ material synthesis for 0.5 hour with 10\% excess resulted in deriving the phase-pure WZ \ce{Mg3WN4}. \textit{Ex situ} PXRD corroborated by SHG measurements shows that this phase crystallizes in a polar $Pmn2_1$ space group with \ce{Mg^{2+}} in a tetrahedral coordination environment. The combination of Rietveld refinements probing the Mg and W site occupancies, EDS spectroscopy analysis of Mg/W ratio for separate particles, and optical measurements of bulk WZ \ce{Mg3WN4} powder suggested complex defect chemistry in WZ \ce{Mg3WN4}. Overall, the results presented in this paper demonstrate \textit{ex situ} synthesizability of the WZ \ce{Mg3WN4} informed by \textit{in situ} synchrotron PXRD and pave a way towards the future property measurements that would assess the viability of this unusual material as a ferroelectric or \ce{Mg^{2+}} ion conductor.

\section{Acknowledgments}
This work was authored at the National Laboratory of the Rockies, for the U.S. Department of Energy (DOE) under Contract No. DE-AC36-08GO28308. 
Primary funding for synthesis and characterization experiments was provided by DOE Basic Energy Sciences, Materials Chemistry program.
Analysis of \textit{in situ} PXRD data at Colorado State University was supported by the National Science Foundation (Grant No.\ DMR-2515517). Measurements of SHG at University of Houston was supported by the Welch Foundation (Grant E-1457).
This research was performed on APS beam time award from the Advanced Photon Source, a U.S. Department of Energy (DOE) Office of Science user facility operated for the DOE Office of Science by Argonne National Laboratory under Contract No. DE-AC02-06CH11357. We would like to thank Dr. Mellie Lemon for assistance with the air-free WDXRF measurement and Dr. Bryon Larson for assistance with the DR measurement. The views expressed in the article do not necessarily represent the views of the DOE or the U.S. Government.

\section{Supporting Information}
\begin{itemize}
\item Additional \textit{in situ} XRD analysis, \textit{ex situ} PXRD experiments on \ce{Mg3WN4}, compositional analysis by EDS and SEM micrograph, ion exchange thermodynamics and salt melting points (PDF).
\item Cation-ordered $Pmn2_1$ structure of \ce{Mg3WN4} (CIF) also deposited in the CCDC
\item Cation-disordered $Fm\bar{3}m$ structure of \ce{MgWN2} (CIF)
\end{itemize}

\begin{tocentry}
\begin{center}
\includegraphics[width = 3.25 in]{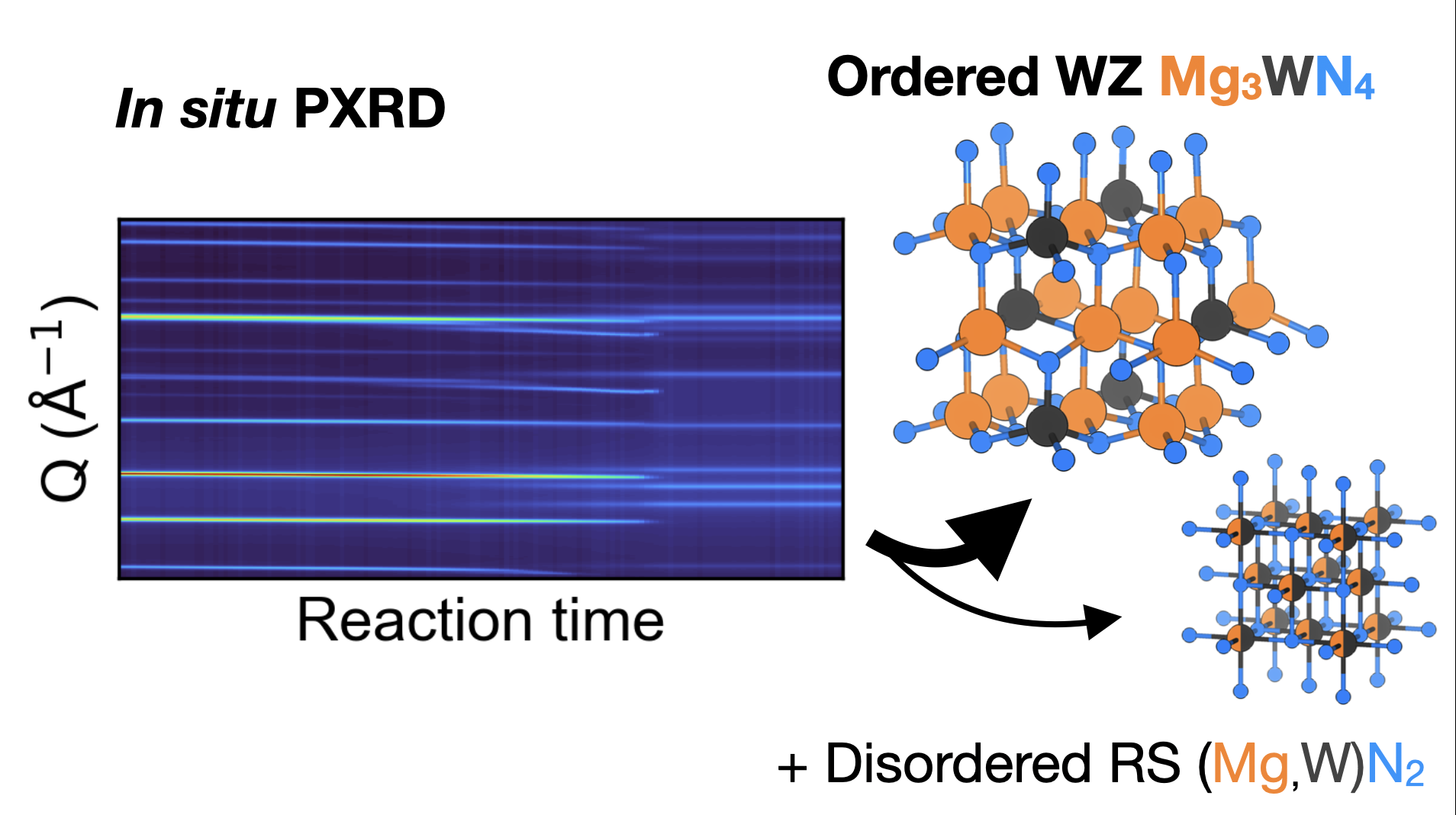}
\end{center}
\end{tocentry}

\providecommand{\latin}[1]{#1}
\makeatletter
\providecommand{\doi}
  {\begingroup\let\do\@makeother\dospecials
  \catcode`\{=1 \catcode`\}=2 \doi@aux}
\providecommand{\doi@aux}[1]{\endgroup\texttt{#1}}
\makeatother
\providecommand*\mcitethebibliography{\thebibliography}
\csname @ifundefined\endcsname{endmcitethebibliography}  {\let\endmcitethebibliography\endthebibliography}{}

\end{document}